\RequirePackage{fix-cm}
\documentclass[twocolumn,epjc3]{svjour3}  
\smartqed  
\RequirePackage{color,graphicx,url}

\journalname{Eur. Phys. J. A}

\usepackage{amsfonts,amsmath,amssymb,bm,xspace}
\usepackage{soul}
\graphicspath{{./figures/}}

\usepackage{hyperref}                

\usepackage{academicons} 
\newcommand{\orcid}[1]{\href{https://orcid.org/#1}{\textcolor[HTML]{A6CE39}{\aiOrcid}}}
\usepackage{xcolor}
\usepackage{hyperref}

\newcommand{\JM}[1]{\textsf{\color{blue}{\textsuperscript{JM}#1}}}
\newcommand{\GG}[1]{\textsf{\color{red}{\textsuperscript{GG}#1}}}

\newcommand{\bat}{\Bigr\rvert}

\newcommand{\sym}{\mathrm{sym}}

\newcommand{\pair}{\mathrm{pair}}
\newcommand{\ffg}{\mathrm{FFG}}

\newcommand{\MM}{\mathrm{MM}}
\newcommand{\MF}{\mathrm{MF}}
\newcommand{\NM}{\mathrm{NM}}
\newcommand{\SM}{\mathrm{SM}}

\newcommand{\nuc}{\mathrm{nuc}}
\newcommand{\tot}{\mathrm{tot}}

\newcommand{\cl}{\mathrm{cl}}
\newcommand{\WS}{\mathrm{WS}}

\newcommand{\crust}{\mathrm{crust}}
\newcommand{\intn}{\mathrm{int}}
\newcommand{\corr}{\mathrm{corr}}

\newcommand{\coul}{\mathrm{Coul}}

\newcommand{\surf}{\mathrm{surf}}
\newcommand{\curv}{\mathrm{curv}}

\usepackage{graphicx}

\hypersetup{%
    pdfsubject=Paper,
    pdfkeywords={nuclear physics} {MBPT} {chiral EFT} {asymmetric nuclear matter}
    unicode=true,
    breaklinks=true,
     colorlinks   = true,
     linkcolor = blue,
     citecolor = blue,
     menucolor = blue,
     urlcolor = blue
}

\begin{document}

\title{Effects of dilute neutron matter on the neutron star crust equation of state}

\author{G. Grams\thanksref{e1,addr1} \and J. Margueron\thanksref{e2,addr2,addr3}}

\thankstext{e1}{e-mail: guilherme.grams@ulb.be}
\thankstext{e2}{e-mail: j.margueron@cnrs.fr}

\institute{Institut d’Astronomie et d’Astrophysique, CP-226, Université Libre de Bruxelles, 1050 Brussels, Belgium \label{addr1}
\and
Institut de Physique des 2 infinis de Lyon, CNRS/IN2P3, Universit\'e de Lyon, Universit\'e Claude Bernard Lyon 1, F-69622 Villeurbanne Cedex, France \label{addr2}
\and
International Research Laboratory on Nuclear Physics and Astrophysics, Michigan State University and CNRS, East Lansing, MI 48824, USA \label{addr3}} 


\maketitle


\abstract{We develop a compressible liquid-drop model to describe the crust of neutron stars for which the role of the nuclear clusters, the neutron gas, and the electrons are clearly identified. The novelty relies on the contribution of the neutron gas, which is qualitatively adjusted to reproduce 'ab initio' predictions in dilute neutron matter. We relate the properties of dilute neutron matter to the ones of neutron stars crust and we compare the mean-field approximation to an improved approach that better describes dilute neutron matter\footnote{This work was a matter of many discussions with Peter Schuck, who had a deep interest in correlated many-body systems and their application in the understanding of the properties of neutron stars.}. The latter is quite sensitive to the unitary limit, a universal feature of Fermi systems having a large value of the scattering length and a small interaction range. While the impact of the accurate description of dilute neutron matter is important in uniform matter (up to 30\% corrections with respect to a mean-field calculations), we find a reduction of this impact in the context of the crust of neutron stars due to the additional matter components (nuclear clusters and electrons). In agreement with our previous works, dilute neutron matter is however a necessary ingredient for accurate predictions of the properties of the crust of neutron stars.}


\section{Introduction}
\label{sec:intro}

Low density neutron matter is predicted to be quite different from ordinary nuclear matter~\cite{Carlson:2003,Gandolfi:2015}. Dilute neutron matter is indeed expected to be close to a universal system, the so-called unitary Fermi gas. The unitary system was introduced by G. Bertsch in 1999 as a toy model for spin-1/2 fermions to describe low density neutron matter~\cite{Baker:1999,Baker:2001} for which the effective range of the interaction $r_e$ is smaller than the inter-particle distance $k_F^{-1}$, which is itself smaller than the absolute value of the scattering length $a_s$: $r_e\ll k_F^{-1}\ll\vert a_s\vert$. In such a system, it is expected that all thermodynamical quantities are proportional to the sole remaining scale given by the Fermi momentum $k_F$~\cite{Carlson:2003}. For instance, the total energy scales with the energy of the free Fermi gas, $E=\xi_s E_{\ffg}$, with $E_{\ffg}=(3/5)E_F$ the energy of the free Fermi gas and $E_F$ is the Fermi energy, $E_F=\hbar^2 k_F^2/(2 m)$. The constant $\xi_s$ is usually called the Bertsch parameter.

Since the properties of this system are universal, cold atom gas at the unitary limit could be used to provide experimental results~\cite{Lobo:2006,Tan:2008,Navon:2010,Kuhnle:2010,Ku:2012,Munekazu:2017,Hiroyuki:2017}: the universal Bertsch parameter $\xi_s$ is measured to be $0.376\pm0.004 $~\cite{Ku:2012}, $0.41\pm0.01$~\cite{Navon:2010}. It has further been suggested that cold atoms may explore the whole regime from unitary to naturalness and could, therefore, provide the neutron matter equation of state (EoS)~\cite{VanWyk:2018,Horikoshi:2019}. A recent review about neutron matter from low to high density can be found in Ref.~\cite{Gandolfi:2015}. 

\begin{figure}[t]
\centering
\includegraphics[scale=0.5]{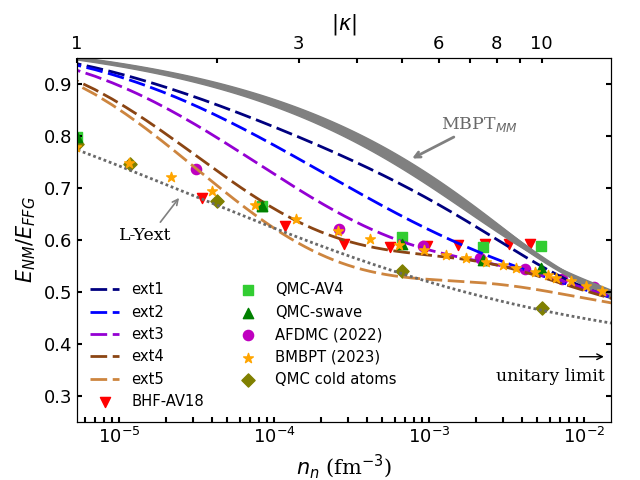}
\caption{Neutron matter energy normalized to the free Fermi gas energy $E_\NM/E_\ffg$ as function of the neutron density $n_n$ in logarithmic scale (bottom axis) and of $\kappa$ (top axis). The gray band represents the uncertainties from the mean-field approximation (meta-model), which is calibrated to reproduce the many-body perturbation theory (MBPT) predictions based on several N2LO chiral-EFT interaction~\cite{Drischler:2016,Drischler:2020hwi}. The dashed lines represent several extensions of the mean-field approximation (ext1 to ext5), which are compared to different 'ab initio' predictions for dilute neutron matter shows with symbols: BHF-AV18~\cite{Vidana:2021} (triangles down), QMC-AV4~\cite{Gezerlis:2010} (squares), QMC-swave~\cite{Gezerlis:2008} (triangles up), AFDMC-2022~\cite{Gandolfi:2022} (full circles) and BMBPT-2023~\cite{Palaniappan:2023} (stars, cutoff $\Lambda = 2 k_F$). The calculation of cold atoms is also shown as QMC cold atoms~\cite{Gezerlis:2008} (diamonds) as well as the asymptotic value for the unitary limit. Finally, the expression~\eqref{eq:Lacroix} is shown (dotted line).} 
\label{fig:fit}
\end{figure}

The actual nuclear interaction is, however, not as simple as its description in the toy model~\cite{Gezerlis:2008,Vidana:2021}: its dimensionless effective range $r_e k_F$ is larger than the cold atom gas one and the nuclear interaction receives contributions from angular momenta $L>0$, which increases in size as the density increases, as illustrated for the respective $S$- and $P$-wave contributions in Ref.~\cite{Gezerlis:2010,Vidana:2021} and reported in Fig.~\ref{fig:fit}, see also Fig.~3 of Ref.~\cite{Gezerlis:2010} and Fig.~1 of Ref.~\cite{Vidana:2021}. Recent predictions for the dilute neutron matter energy based on different treatments of the many-body correlations~\cite{Gezerlis:2008,Gezerlis:2010,Vidana:2021,Gandolfi:2022,Palaniappan:2023} are shown in Fig.~\ref{fig:fit} as well as quantum Monte-Carlo (QMC) calculation for cold atoms~\cite{Gezerlis:2008}. It is also clear in Fig.~\ref{fig:fit} that the predictions for dilute neutron matter and cold atom gas are close at very low density, but differences appear and increase as the density increases, for reasons briefly detailed here before.
Another remark is that the ratio $E_\NM/E_\ffg$ is close to being constant in dilute neutron matter for densities in the range $10^{-4}$ to about $10^{-2}$~fm$^{-3}$, but the ratio is about 0.6 instead of the value of the Bertsch constant for the unitary limit.

In a recent study, we have shown the importance of the neutron gas component for the understanding of the neutron stars' crust properties by comparing different types of Hamiltonians~\cite {Grams2022b}. A first set of Hamiltonians was based on chiral EFT interactions and a second set was aggregating predictions based on various Skyrme interactions. All these Hamiltonians describe matter at the mean field approximation. We found that the small (large) dispersion in the prediction of neutron matter properties of the first (second) set could be related to the small (large) dispersion in the predictions of the neutron star crust equation of state. Building upon this result illustrating the important role of the neutron gas to the neutron star crust properties, we investigate in this paper the additional effect induced by the correlations in dilute neutron gas compared to the mean-field predictions. In the following, we describe our treatment of these additional correlations and implement them in the modeling of the neutron star crust equation of state.

We further elaborate on the properties of Fermi systems at low density in Sec.~\ref{sec:unitary}. The formalism and our results for uniform neutron matter are presented in Sec.~\ref{sec:neutronmatter}, then non-uniform modeling of matter in the crust of neutron stars is described in Sec.~\ref{sec:crust} together with our results. Conclusions are presented in Sec.~\ref{sec:conclusions}.

\section{Low density Fermi systems}
\label{sec:unitary}

Considering a N-particle quantum-mechanical system interacting via two-body hard sphere potentials at extremely low densities Lee and Yang have obtained the following analytical expression for the energy~\cite{Lee:1957},
\begin{equation}
\frac{E_{L-Y}}{E_{\ffg}}=1+\frac{10}{9\pi}\kappa+\frac{4}{21\pi^2}(11-2\log 2)\kappa^2 + o(\kappa^2) \, ,
\label{eq:LY}
\end{equation}
as a function of the parameter $\kappa$, which is defined as $\kappa=a_s k_{F}$ for cold atom gas and $\kappa=a_{nn} k_{F_n}$ for dilute neutron matter. By construction, this expression is valid only for $\vert\kappa\vert\ll 1$. 

The following extension of Eq.~\eqref{eq:LY} was suggested in Ref.~\cite{Lacroix:2016},
\begin{equation}
\frac{E_{L-Y}^{ext}}{E_{\ffg}}=1+\frac{10}{9\pi}\kappa
\left[1-\frac{10}{9\pi}\frac{\kappa}{1-\xi(k_F r_e)}\right]^{-1}\, ,
\label{eq:Lacroix}
\end{equation}
where the function $\xi(x)$ is defined as~\cite{Lacroix:2016}: 
\begin{equation}
\xi(x) = 1-\frac{(1-\xi_0)^2}{1-\xi_0+x\eta_e}
\end{equation}
with $\xi_0$ and $\eta_e$ two parameters adjusted to reproduce the universal limit of a unitary Fermi gas. Note that $\xi(x=0)=\xi_0$. It was suggested in Ref.~\cite{Vidana:2021} to take $\eta_e=0$ and $\xi_0=0.37$ to reproduce cold atom gas predictions for $\vert\kappa\vert<1$ and $\vert\kappa\vert>1$, see Fig.~\ref{fig:fit} for instance where the relation~\eqref{eq:Lacroix} (L-Yext) is shown in dotted line and reproduce very well the cold atom gas QMC calculations.

Dilute neutron matter is known to probe the cross-over region located in-between the weak (BCS) and the strong (BEC) regimes~\cite{Matsuo:2006,Margueron:2007}, illustrating the importance of the residual correlations. These correlations induce super-fluidity for instance~\cite{Cao:2006}, and the BCS formalism~\cite{BCS:1957} could be applied in the two extreme limits of weak and strong couplings, as shown by Nozi\`eres and Schmitt-Rink~\cite{Nozieres:1985}. These correlations are also important for the understanding of the link between neutron-rich nuclei located at the neutron drip line~\cite{Pastore:2013} and cold atom gases~\cite{Pastore:2014}.

Note that the effect of pairing in Eq.~\eqref{eq:LY} is ignored since it is expected to be small at extremely low density, $E_{\pair}\approx \exp(\pi/\kappa)$~\cite{DeGennes:1966}. While this is correct for
$\vert\kappa\vert\ll 1$ and $\kappa<0$, it may not hold true for larger values of $\vert\kappa\vert$. In the present study, we extend the Fermi gas description given in Eqs.~\eqref{eq:LY} by considering the effect of correlations beyond the mean field approximation in dilute neutron matter.

To do so, the meta-model~\cite{Margueron:2018b,Somasundaram2021} (MM) constrained by chiral EFT Hamiltonians~\cite{Somasundaram2021,Grams2022b} is taken as a reference for the mean field description of dilute matter and is complemented by an additional term mimicking correlations beyond mean-field. This extension is calibrated so that it can be simply implemented in modeling the crust of neutron stars~\cite{Grams2021,Grams2022a,Grams2022b}, see Sec.~\ref{sec:crust}. 

\section{Dilute neutron matter}
\label{sec:neutronmatter}

\begin{table}[t]
\begin{center}
\caption{Dimensionless parameters of the field $\Delta_n$ described by Eq.~\eqref{eq:fieldn}.}
\label{table:pairparam}
\begin{tabular}{c|ccccc}
\hline\noalign{\smallskip}
    & ext1 & ext2 & ext3 & ext4  & ext5  \\
\hline\noalign{\smallskip}
a ($\times 10^{-2}$) &  -1.615 & -0.0322 & 33.59    & 95.94 & 113.1 \\
b ($\times 10^{-6}$) &  0.6784 & 6.589   & 0.0335    & 5.497 & 0.171 \\
c ($\times 10^{-2}$) &  10.95  & 0.704   & 1.236    & 7.420 & 8.137 \\
d                    &  1.107  & 2.144   & 2.015    & 1.336 & 1.239 \\
\noalign{\smallskip}\hline
\end{tabular}
\end{center}
\end{table}

We consider spin-saturated dilute neutron matter as calculated in the following references~\cite{Gezerlis:2008,Gezerlis:2010,Vidana:2021,Gandolfi:2022,Palaniappan:2023} and represented in Fig.~\ref{fig:fit}. Since these calculations slightly differ from each other, and we want to explore the impact of extending the mean-field prediction (gray band) step-by-step, we have generated a set of extensions (ext1 to ext5) that gradually depart from the mean-field reference calculation, see Fig.~\ref{fig:fit}. For this reason, extensions ext1 to ext5 are reproducing only qualitatively the 'ab initio' predictions for dilute neutron matter. Note that since pairing is not considered in the chiral EFT predictions from Ref.~\cite{Drischler:2016,Drischler:2020hwi}, we employ these results to adjust a set of meta-models reproducing the mean-field approximation, see Ref.~\cite{Somasundaram2021}, which lies in the gray band shown in Fig.~\ref{fig:fit}.

\begin{figure}[t]
\centering
\includegraphics[scale=0.5]{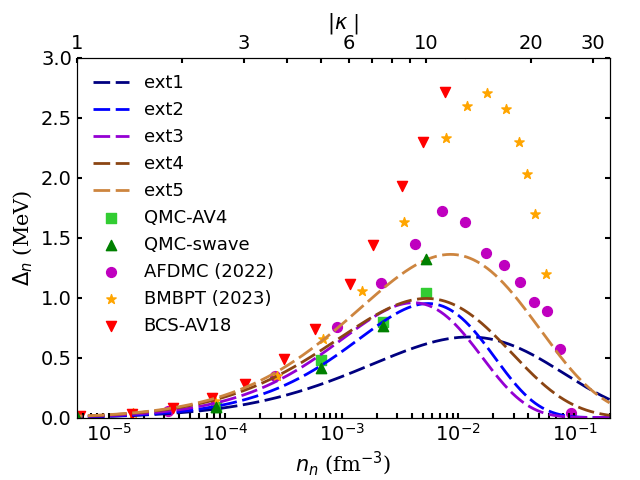}\\
\includegraphics[scale=0.5]{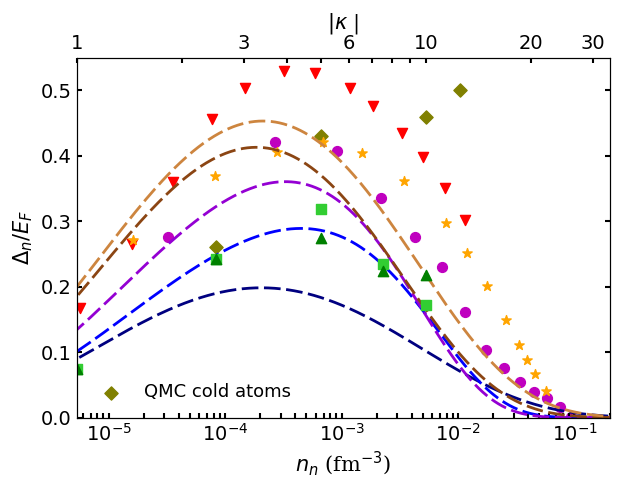}
\caption{Field $\Delta_n$ and 'ab initio' pairing gaps (top panel) and same quantities normalized by the Fermi energy $E_F$ (bottom panel) w.r.t to the neutron density $n_n$ (lower x-axis) and variable $\kappa$ (upper x-axis) in neutron matter. See Fig.~\ref{fig:fit} for more details on the legend.}
\label{fig:gapn}
\end{figure}

The energy density in dilute neutron matter is defined as $\epsilon_\tot \equiv \epsilon_\mathrm{mass} + \epsilon_\intn$, where $\epsilon_\mathrm{mass}$ is the rest mass energy density and the internal energy density $\epsilon_\intn$ is approximated by the following expression:
\begin{equation}
\epsilon_\intn \approx \epsilon_\MF + \epsilon_\corr \, ,
\label{eq:totintnrj}
\end{equation}
where $\epsilon_\MF$ is the mean-field energy density given by the meta-model $\epsilon_\MF=\epsilon_\MM$, considering the uncertainty band shown in Fig.~\ref{fig:fit}, and $\epsilon_\corr$ is a correction energy capturing correlations beyond the mean field approximation. Since the uncertainty in the meta-model is small, we select one of them (H2$_{\MM}$) for the rest of the discussion.
Results obtained with H2$_{\MM}$ functional provide, therefore, reference calculations with respect to which extensions are compared in order to evaluate the impact of the correlations in dilute neutron matter.

\begin{figure*}[t]
\centering
\includegraphics[scale=0.6]{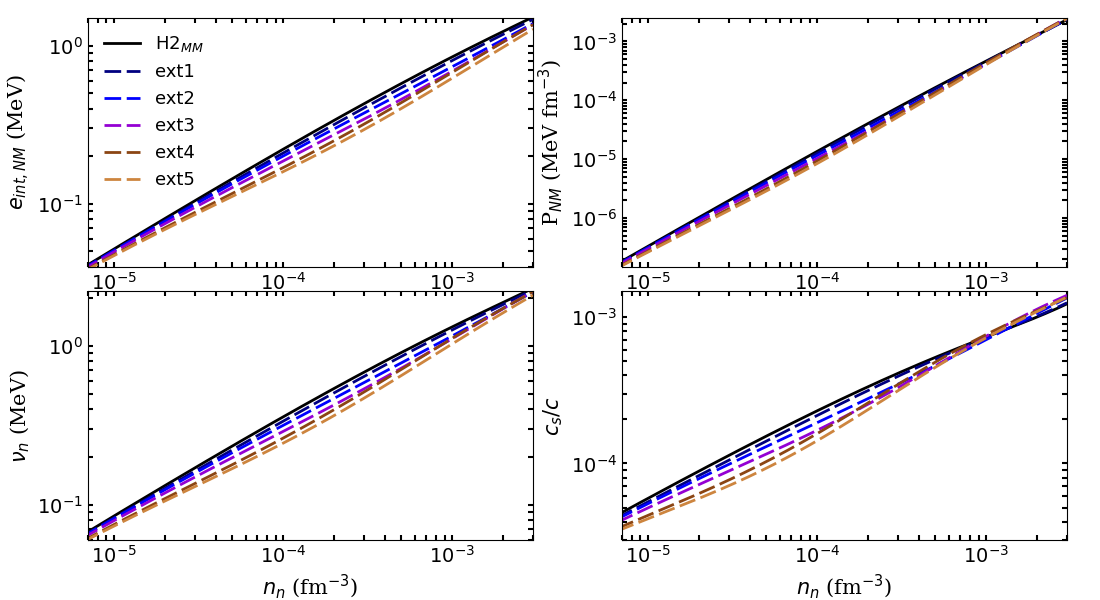}
\caption{Neutron matter internal energy per particle $e_{\intn,\NM}$, pressure $P_\NM$, internal chemical potential $\nu_n$, and sound speed $c_s/c$ ($c$ is the speed of light in vacuum) w.r.t the neutron density $n_n$ for H2$_{\MM}$ (solid line) and the extensions ext1 to ext5 (dashed lines).}
\label{fig:thermoNM}
\end{figure*}

The correlation energy $\epsilon_\corr$ is defined in the following way~\cite{DeGennes:1966,Yang1971}
\begin{equation}
\epsilon_\corr = - \frac 1 2 N_{0n} \big(\Delta_n(n_n)\big)^2  \, , 
\label{eq:corre}
\end{equation}
where appear the density of state $N_{0n}=m^*_n k_{F_n}/(\pi^2\hbar^2)$, the in-medium mass $m^*_n$, the Fermi momentum $k_{F_n}$ defined as $n_n=k_{F_n}^3/(3\pi^2)$, and a field $\Delta_n(n_{n})$ which is adjusted to the 'ab initio' predictions for dilute neutron matter. 

In the following, the field $\Delta_n(n_{n})$ matches with the pairing gap at extremely low density, see the discussion hereafter, but for higher densities, it should not be interpreted as the pairing gap of dilute neutron matter. It mimics the effect of the many-body correlations beyond the mean field approximation in a way that makes it convenient to model the crust of neutron stars. Since the many-body correlations are expected to be large only in dilute neutron matter, the field $\Delta_n(n_{n})$ is parameterized in the following way,
\begin{equation}
\Delta_n (n_{n}) = \Delta^{\rm lowk}_n 
[1+a|\kappa|^b]
\exp{[-c |\kappa|^d]} \, ,
\label{eq:fieldn}
\end{equation}
where the exponential term strongly reduces the value of the field $\Delta_n(n_{n})$ as the density increases, i.e., for $k_{F_n}\gg c^{1/d}/\vert a_{nn} \vert$~fm$^{-1}$.

The values of the four parameters $a$, $b$, $c$, $d$ are given in Tab.~\ref{table:pairparam} for the five extensions (ext1-ext5) represented in Fig.~\ref{fig:fit}. 
These values are arbitrary and they are fixed such that the total internal energy~\eqref{eq:totintnrj} gradually evolves from the MBPT prediction to the 'ab initio' prediction for dilute neutron matter. The parameters $a$ and $b$ induce a polynomial correction in $\vert\kappa\vert$ to the field $ \Delta^{\rm lowk}_n$, which represents the very low density limit of the field $\Delta^{\rm lowk}_n$. This limit is dominated by pairing correlations, and therefore $\Delta^{\rm lowk}_n$ is taken to reproduce the very low-density pairing gap. This value has been derived by Gor’kov and Melik-Barkhudarov~\cite{Gorkov:1961} and is fixed by the following analytical expression:
\begin{equation}
\Delta_n^\mathrm{low k} = \frac{8}{e^2} E_{F} \exp\left(\frac{\pi}{2\kappa} -\frac{\pi \bar{c}}{2} \right) \, ,
\label{eq:gappol}
\end{equation}
with the Fermi energy $E_{F}$ and the constant $\bar{c}=\frac{2}{3\pi}(1+2\log 2)\approx 0.506$. Note that Eq.~\eqref{eq:gappol} provides a value of the pairing gap at very low density which incorporates the effects of polarization (exchange of excitations in the medium), i.e. a reduction of the low coupling pairing field, also referred as the BCS pairing gap~\cite{Heiselberg:2000,Vidana:2021}, by a factor of about $\approx 2$.

Note that our results are not bounded to the specific form chosen for the correlation energy~\eqref{eq:corre} with the field~\eqref{eq:fieldn}. A different choice would lead to similar conclusions for dilute neutron matter.

In Fig.~\ref{fig:gapn}, we represent the field $\Delta_n(n_{n})$ obtained for the extensions ext1-ext5 shown in Fig.~\ref{fig:fit} (dashed lines) as well as the pairing gaps obtained from the 'ab initio' calculations given in Refs.~\cite{Gezerlis:2008,Gezerlis:2010,Vidana:2021,Gandolfi:2022,Palaniappan:2023} (symbols). While these quantities are different in nature, they are shown to be quite similar in absolute value. The absolute values of the field $\Delta_n(n_{n})$ and of the pairing gap (top panel) are scaled w.r.t. the Fermi energy $E_F$ (bottom panel). The latter reflects more realistically the region of density where the correlations beyond the mean field approximation are expected to be stronger: in the density range between $10^{-5}$ and $10^{-3}$~fm$^{-3}$. 
The peak shown on the top panel at about $\approx 10^{-2}$~fm$^{-3}$ is not located in the most important region for dilute neutron matter. Another interesting feature from Fig.~\ref{fig:gapn} is that the two quantities, the field $\Delta_n$ and the pairing gaps extracted from 'ab initio' calculations are quite similar in absolute value. This was not intentional, except at extremely low density where we impose a matching between these two quantities. It however tend to show that even for the densities of interest in this study, i.e., $10^{-4}$ to $10^{-2}$, pairing energies may be the dominant contribution to the correlations beyond the mean field, but of course not the sole one.

\begin{figure}[t]
\centering
\includegraphics[scale=0.45]{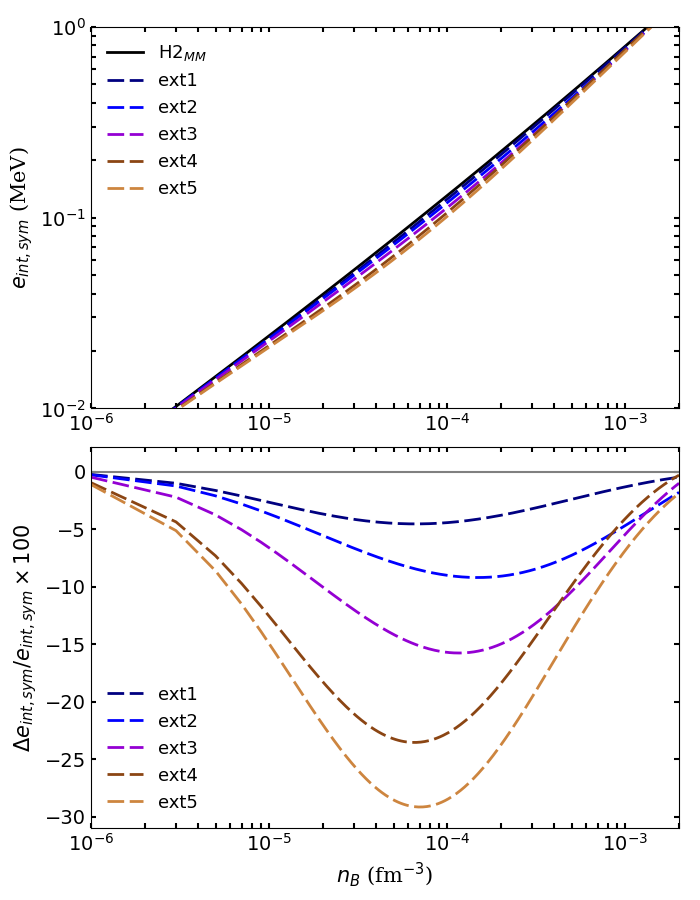}
\caption{Internal symmetry energy $e_{\intn,\sym} = e_{\intn,\NM} - e_{\intn,\SM}$ with respect to the baryon density $n_B$ in homogeneous matter. The relative difference $\Delta e_{\intn,\sym}/e_{\intn,\sym}$ reflects the difference between the prediction within the extended description of the dilute neutron matter and the prediction based on the mean field approximation. The denominator is given by the mean field approximation.} 
\label{fig:esym}
\end{figure}

We represent in Fig.~\ref{fig:thermoNM} some thermodynamical quantities in dilute neutron matter predicted by the extensions (dashed lines) to the meta-model shown in Fig.~\ref{fig:fit}: the internal energy per particle, the pressure, the internal chemical potential $\nu_q = (\partial \epsilon_{\tot}) /\partial n_q|_{n_{\Bar{q}}}  - m_{q}c^{2}$, where $\bar{q}$ is $n(p)$ for $q=p(n)$, and the sound speed. Note that we also represent the prediction of the mean field model H2$_{\MM}$ without extensions (solid line). The reduction of the energy per particle in the extended models compared to the mean field model implies a softening of the energy for densities $n_n<10^{-3}$~fm$^{-3}$ and a hardening for $n_n>10^{-3}$~fm$^{-3}$, which is reflected in the derivatives of the energy per particle: the pressure, the internal chemical potential and the sound speed (second derivative). 

\begin{figure}[t]
\centering
\includegraphics[scale=0.45]{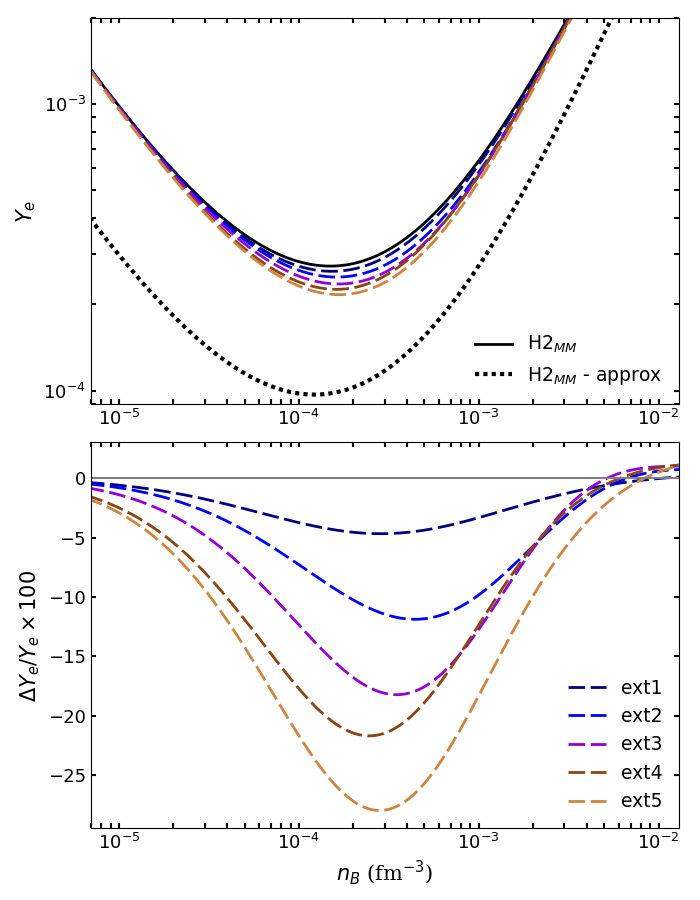}
\caption{Electron fraction with respect to the baryon density $n_B$ in $\beta$-equilibrated homogeneous matter. The relative difference $\Delta Y_e/Y_e$ is calculated similarly to the one given in Fig.~\ref{fig:esym}.
} 
\label{fig:yeappr}
\end{figure}

The sound speed $c_s$ is defined from the derivative of the total pressure $p$ with respect to the energy-density $\epsilon_\tot$,
\begin{eqnarray}
(c_s/c)^2 &\equiv& \frac{d p}{d \epsilon_\tot} 
 = \frac{K(n)}{h(n)} \, , 
\end{eqnarray}
where $c$ is the speed of light in vacuum, $K(n)=\partial p/\partial n$ is the matter incompressibility and $h(n) = mc^2 + e(n) + p(n)/n$ is the enthalpy. Since $h\approx mc^2$ at low density, the sound speed is mostly impacted by the derivative of the pressure, as shown in Fig.~\ref{fig:thermoNM}.

The internal symmetry energy is defined as the difference between the neutron matter internal energy per particle $e_{\intn,\NM}$ and the same value in symmetric matter $e_{\intn,\SM}$~\cite{Somasundaram2021}:
\begin{equation}
e_{\intn,\sym}(n) = e_{\intn,\NM}(n) - e_{\intn,\SM}(n) \, .
\end{equation}
Note that the neutron correlation term is included in both neutron and symmetric matter. The internal symmetry energy is shown in Fig.~\ref{fig:esym}.
The impact of the extensions can be important, modifying the internal symmetry energy by up to 30\% at low density for $n<10^{-3}$~fm$^{-3}$. 

In Fig.~\ref{fig:yeappr} the electron fraction in homogeneous $\beta$-equilibrated matter is shown. As expected from the corrections to the symmetry energy, the electron fraction is also modified by a large amount, up to about 25\% for $n<10^{-3}$~fm$^{-3}$.

The electron fraction is obtained from the solution of the $\beta$-equilibrium equation:
\begin{equation}
\nu_{n} = \nu_{p} + \nu_e + \Delta mc^2 ,
\label{eq:betaUM}
\end{equation}
with
$ \Delta mc^2=(m_p+m_e-m_n)c^2\approx -0.78$~MeV.
At $T=0$ and approximating $\nu_{n}-\nu_{p}\approx 2\delta e_{\intn,\sym(n)}$, we obtain the following expression for relativistic electrons:
\begin{equation}
\hbar\left( 3\pi^2 Y_e n\right)^{1/3} \approx  4 ( 1-2 Y_e) e_{\intn,\sym}(n) -\Delta mc^2\, .
\label{eq:betaUM2}
\end{equation}
At low density, since $Y_e\ll 1$, we obtain from Eq.~\eqref{eq:betaUM2} the following approximation:
\begin{equation}
Y_e \approx \frac{1}{3\pi^2 \hbar^3}\frac{(4e_{\intn,\sym}-\Delta mc^2)^3}{n} \, ,
\end{equation}
which is represented in Fig.~\ref{fig:yeappr} (dotted line).

In summary, the 'ab initio' predictions for dilute neutron matter, as captured by the different extensions we study, are important and can lead to substantial corrections to the mean-field predictions. The realistic case of the crust of a neutron star shall however include, in addition, the contributions from the nuclear clusters and the leptons. This is the aim of the next section.

\section{Neutron star crust}
\label{sec:crust}

The cold-catalyzed crust of neutron stars is a physical system, different from uniform matter presented in Sec.~\ref{sec:neutronmatter}, that exists in nature. It is a non-uniform system containing nuclear clusters and electrons in addition to dilute nuclear matter. The respective weights of these different contributions are given by the modeling presented in this section. Note that since the uniformity of matter is broken by the sole presence of the nuclear clusters, these clusters are sometimes referred to as impurities.

\begin{figure}[t]
\centering
\includegraphics[scale=0.25]{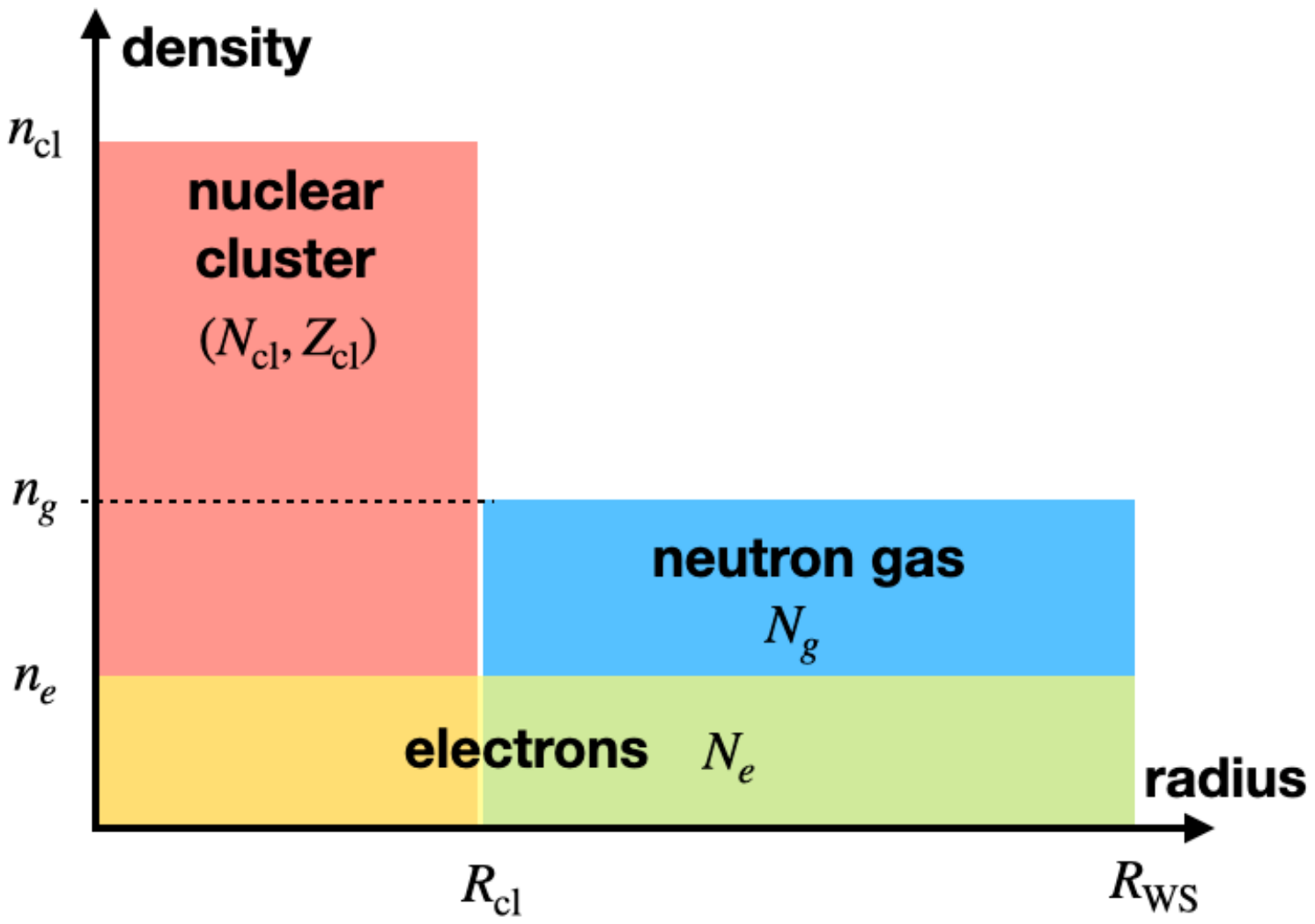}
\caption{Schematic description of the $r$-representation of the spherical Wigner-Seitz cell.}
\label{fig:WS}
\end{figure}

In this section, we consider the dilute neutron matter description presented in Sec.~\ref{sec:neutronmatter} into a global modeling of the neutron star crust. The properties of the crust can be obtained from its reduction to a unit cell, the Wigner-Seitz cell of volume $V_\WS$, surrounding a single nuclear cluster of volume $V_\cl$. Matter in a mesoscopic region of the crust is assumed to be obtained by the pure replication of this geometry. We adopt the $r$-representation of the spherical Wigner-Seitz cell~\cite{Papakonstantinou:2013} schematically shown in Fig.~\ref{fig:WS}: the neutron gas is treated as a uniform system occupying the total volume $V_\WS$ (with radius $R_\WS$) minus the volume occupied by the cluster $V_\cl$, while electrons fill the entire volume $V_\WS$ and the nuclear clusters occupy the volume $V_\cl$ delimited by the radial coordinate $R_\cl$. 

In terms of densities, the total density $n_\tot$ in a mesoscopic region of the crust receives contributions from the nuclear clusters $n_\cl$, the neutron gas $n_g$, and the electrons $n_e$ and can be written as $n_\tot = n_B+n_e=(1+Y_e)n_B$ with the baryon contribution
\begin{equation}
n_B = u n_\cl + (1-u) n_g\, ,
\label{eq:ntot}
\end{equation}
and the lepton contribution $n_e=Y_e n_B$, assuming that only electrons are present in the crust (muons appear at higher densities). In Eq.~\eqref{eq:ntot}, the quantity $u$ is the volume fraction, which can be expressed as the ratio between the cluster volume $V_\cl$ over the Wigner-Seitz volume $V_\WS$,
\begin{equation}
u = \frac{V_\cl}{V_\WS} = \frac{2n_e}{(1-I_\cl)n_\cl} \, ,
\label{eq:uvol}
\end{equation}
where we have used the electro-neutrality condition specifying that, in a Wigner-Seitz cell, the number of positive charges ($Z_\cl$ protons) equalizes the number of negative charges ($N_e$ electrons): $Z=N_e$. In Eq.~\eqref{eq:uvol} the variable $I_\cl$ represents for the isospin asymmetry in the nuclear cluster $I_\cl=(N_\cl-Z_\cl)/A_\cl$ and $n_\cl$ is the cluster density. 

\begin{figure*}[t]
\centering
\includegraphics[scale=0.6]{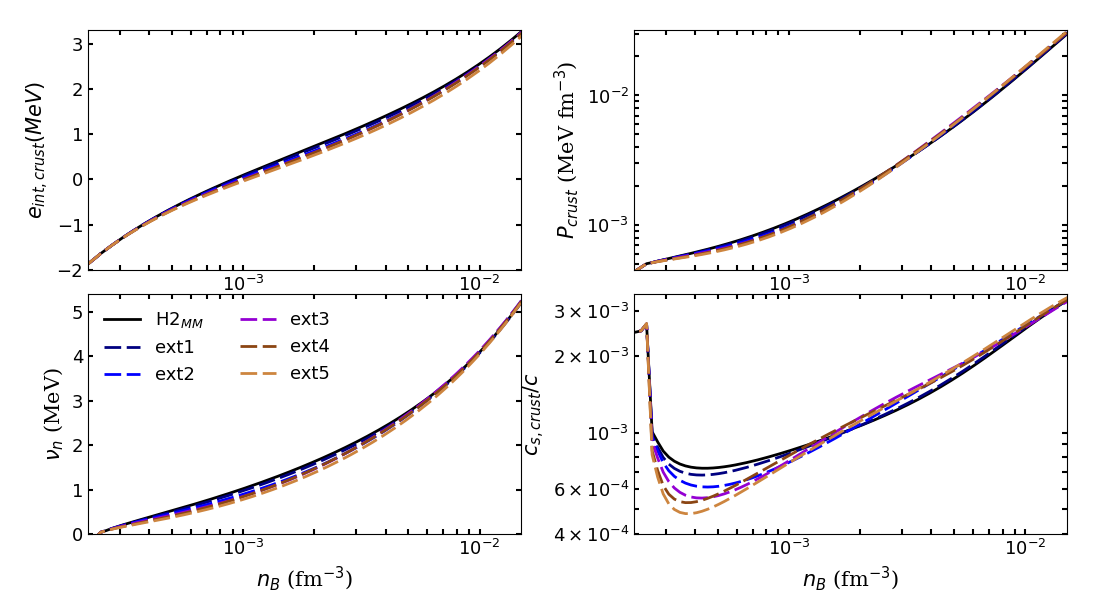}
\caption{Crust internal energy per particle $e_{\intn,\crust}$, pressure $P_\crust$, internal neutron chemical potential $\nu_n$, and sound speed $c_{s,\crust}/c$ w.r.t the baryon density $n_B$ for H2$_{\MM}$ (solid line) and the extended models ext1 to ext5 (dashed lines).}
\label{fig:thermo}
\end{figure*}

\begin{figure*}[t]
\centering
\includegraphics[scale=0.6]{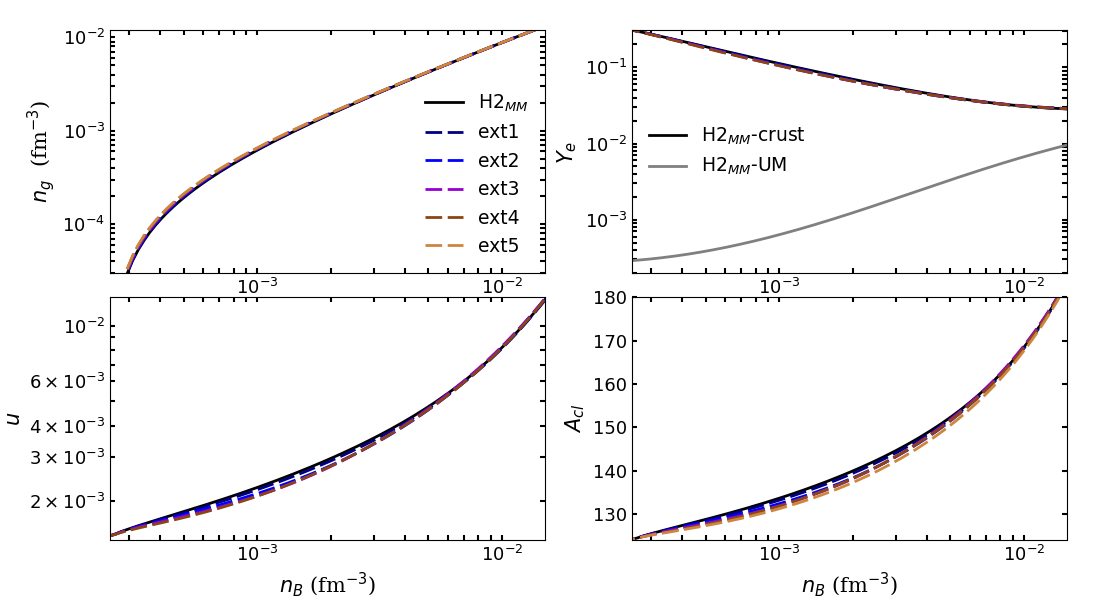}
\caption{Crust properties: neutron gas density $n_g$, electron fraction $Y_e$, volume fraction $u$ and cluster particle number $A_\cl$. We adopt the same notations as in Fig.~\ref{fig:thermo}.}
\label{fig:densities}
\end{figure*}

\begin{figure}[t]
\centering
\includegraphics[scale=0.45]{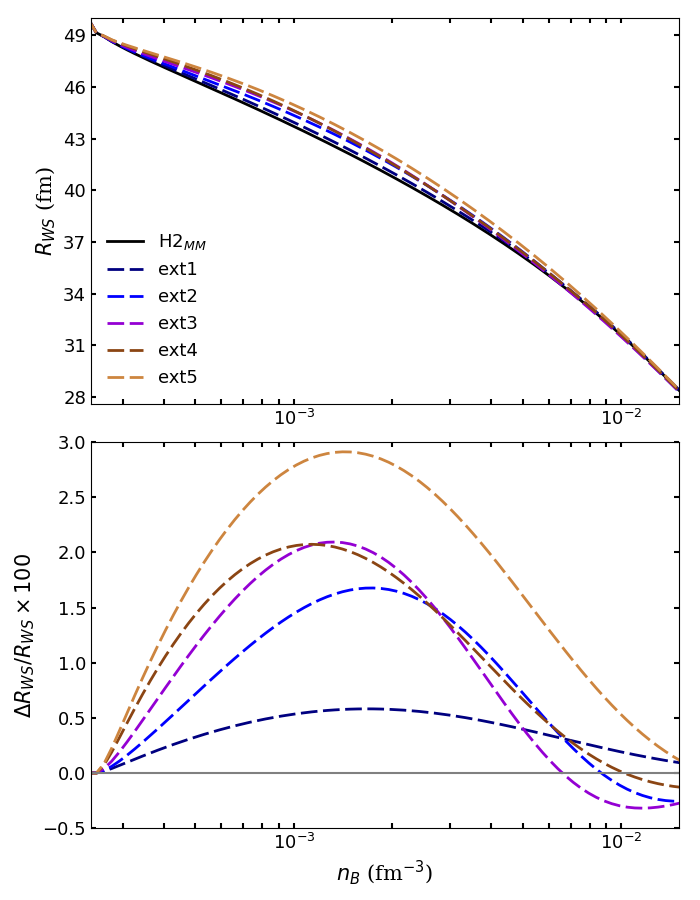}
\caption{Radius of the Wigner-Seitz cell in the neutron star inner-crust for H2$_{\MM}$ (solid black) and the five extensions (dashed lines). The relative difference $\Delta R_\WS/R_\WS$ is calculated similarly to the one in Fig.~\ref{fig:esym}. We adopt the same notations as in Fig.~\ref{fig:thermo}.}
\label{fig:rws}
\end{figure}

The internal energy density in a mesoscopic region of the crust receives contributions from the different components and can be expressed as
\begin{equation}
\epsilon_{\intn,\crust} = \epsilon_\cl+\epsilon_g+\epsilon_e \, ,
\end{equation}
where $\epsilon_\cl$ is the contribution of the nuclear cluster (including the contribution from the interface between the cluster and the neutron gas), $\epsilon_g$ is the neutron gas contribution (given by the modeling of dilute matter in Sec.~\ref{sec:neutronmatter}) and $\epsilon_e$ is the relativistic electron contribution. The cold-catalyzed crust is approximated within the compressible liquid drop model (CLDM) with Coulomb (direct and exchange), surface and curvature terms, as detailed in Refs.~\cite{Grams2021,Grams2022a,Grams2022b}. We employ the more advanced FS4 model~\cite{Grams2022a}, for which the following set of non-linear and coupled stability equations are solved:
\begin{eqnarray}
2E_\coul &=& E_\surf + 2 E_\curv \, , \label{eq:virial}\\
P_\cl &=& P_g \, , \label{eq:meca}\\
\mu_{\cl, n} &=& \mu_g \, , \label{eq:chemical}\\
\mu_{\cl, n} &=& \mu_{\cl, p} + \mu_e + \Delta mc^2 \, , \label{eq:beta}\\
\mu_B &=& \mu_g \nonumber \\
&&+ \frac{2n_e}{n_\cl A_\cl (1-I_\cl)-2n_e}\frac{\partial E_\surf}{\partial n_g}\bat_{A_\cl, I_\cl, n_\cl}
\label{eq:pot} \, ,
\end{eqnarray}
where the quantities $E$ in Eq.~\eqref{eq:virial} stand for total energies in the Wigner-Seitz cell. Because of the non-linear and coupled nature of these equations, it is difficult to anticipate quantitatively the effect of the correlations in dilute neutron matter from the solution of these equations without correlations.
The crust pressure and chemical potentials are defined as,
\begin{eqnarray}
P_\cl &\equiv& n_\cl^2 \frac{\partial E_\cl/A_\cl}{\partial n_\cl}\bat_{A_\cl,I_\cl} \label{eq:Pcl} \, , \\
P_g &\equiv& -\epsilon_g + n_g \mu_g \label{eq:Pg}\, ,\\
\mu_{\cl, q} &\equiv& \mu_{\nuc, q} + \frac{P_g}{n_B} \label{eq:mucl}\, , \\
\mu_e &\equiv& \frac{\partial E_e}{\partial N_e}\bat_{N_\cl, Z_\cl} +\frac{2 n_e}{(1-I_\cl)A_\cl} \frac{\partial E_\coul}{\partial n_e} \label{eq:mue}\, , 
\end{eqnarray}
with
\begin{equation}
\mu_{\nuc, n} \equiv \frac{\partial E_\nuc}{\partial N_\cl}\bat_{Z_\cl, N_e} \, \hbox{and} \, \mu_{\nuc, p} \equiv \frac{\partial E_\nuc}{\partial Z_\cl}\bat_{N_\cl, N_e} \, .
\end{equation}
Note that the cluster chemical potentials~\eqref{eq:mucl} are modified by the neutron gas in the inner crust ($P_g\ne 0$).


Results are shown in Fig.~\ref{fig:thermo}, where the same thermodynamical quantities as in Fig.~\ref{fig:thermoNM} are plotted for the crust. The comparison of Figs.~\ref{fig:thermoNM} and Fig.~\ref{fig:thermo} shows the additional contribution of the nuclear clusters and the electrons. The baryon densities in Fig.~\ref{fig:thermo} are chosen to span over the inner-crust, where the neutron gas density given in Fig.~\ref{fig:thermoNM} goes from zero (at the outer-crust/inner-crust transition) up to about $n_B$ in the densest layers. Note also that since the contribution of the electrons to the density is always negligible, we have $n_{\tot}\approx n_B$.
The dispersion among the extensions applied to the crust and shown in Fig.~\ref{fig:thermo} are smaller than the one observed in Fig.~\ref{fig:thermoNM}. This is due to the additional contribution from the clusters and the electrons which reduces the impact of the neutron gas to the properties of the neutron star crust. The rest of our results follow the same tendency.

In the following figures, Figs.~\ref{fig:densities} and \ref{fig:rws}, we represent additional properties of the crust such as the neutron gas density $n_g$, electron fraction $Y_e$, volume fraction $u$, the cluster particle number $A_\cl$ and the Wigner-Seitz cell radius $R_\WS$. While the neutron gas density and the electron fraction are almost not impacted by the energy correction in dilute neutron matter, the other properties like the volume fraction, the number of nucleons in the cluster, and the Wigner-Seitz radius are more impacted.

The neutron gas density is quite insensitive to the improved modeling of dilute neutron matter because, for the densities where it matters, the neutron gas density is almost identical to the baryon density $n_B$. The electron fraction is also weakly impacted for a different reason: Once the modeling of dilute neutron matter is well constrained by 'ab initio' predictions, the electron fraction is dominantly driven by the presence of clusters in the crust of neutron stars. In order to quantify the influence of the clusters on $Y_e$, the electron fraction of a uniform system without nuclear clusters and at $\beta$-equilibrium (H2$_\MM$-UM) is shown in Fig.~\ref{fig:densities}. This quantity was represented previously in Fig.~\ref{fig:yeappr} as a function of the neutron density while it is now shown as a function of $n_B$, where the relation $n_n(n_B)$ is given by the solution of the CLDM equations. The difference observed between the curve H2$_\MM$-UM and the set of curves H2$_\MM$-crust illustrate the contribution of the nuclear clusters to $Y_e$. This explains why the improved modeling of dilute neutron matter very weakly impacts the quantity $Y_e$. We remind however that in this figure, we compare two models which reproduce the same dilute neutron matter prediction. If this constraint is not satisfied, we recover our previous result shown in Ref.~\cite{Grams2022b}: $Y_e$ is strongly influenced by the Hamiltonians describing neutron matter.

Although the effect of the improved modeling of dilute neutron matter is more visible for $u$, $A_\cl$ and $R_\WS$ shown in Figs.~\ref{fig:densities} and \ref{fig:rws}, the relative correction remains smaller than 10\%. We note for instance a 2-3$\%$ correction at maximum for $R_\WS$ shown in Fig.~\ref{fig:rws}. In the crust of neutron stars, the contribution of the improved modeling of dilute neutron matter does not seem to be a dominant effect for the results we have presented in our analysis. It is however not negligible and contributes to modifying the properties of the crust up to less than 10\%. Even if the effect of correlations is small, considering the important contribution from the mean-field to the prediction of the equation of state in neutron star crust~\cite{Grams2022b}, 'Ab initio' predictions for dilute neutron matter provide important constraints. 

A connection between dilute neutron matter and the crust of neutron stars has recently been performed employing a model of phase coexistence of dense neutron-rich nuclear clusters and dilute neutron gas~\cite{Gupta:2023}. In such a model, the low-density component is obtained from dilute neutron matter, and the high-density component from an equation of state for uniform matter obtained from different Skyrme interactions. This model allows one to connect together different functionals specifically defined to describe dilute neutron matter or nuclear clusters. It however neglects the effect of Coulomb as well as of the interface between the nuclear clusters and the neutron gas. Even though the formalism is different between our work and the one presented in Ref.~\cite{Gupta:2023}, results are qualitatively in agreement: the effect of the improved description of dilute matter reduces the volume fraction $u$, the electron fraction $Y_e$, the pressure $P_\crust$ at low density. The pressure (volume fraction) is however found to be a factor of about two larger (lower) in Ref.~\cite{Gupta:2023} compared to the one we calculate.

\begin{figure*}[t]
\centering
\includegraphics[scale=0.6]{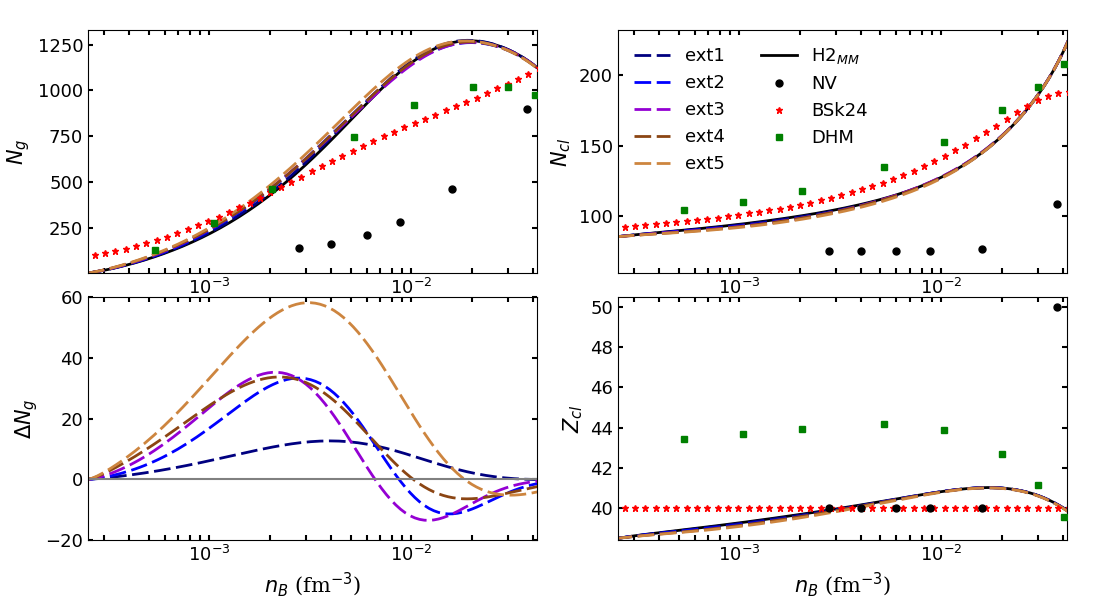}
\caption{Nuclear cluster properties in Wigner-Seitz cells: number of neutrons in the gas $N_g$, in the cluster $N_\cl$ and number of protons in the cluster $Z_\cl$. We show predictions based on H2$_{\MM}$ alone (solid line) and the extensions ext1 to ext5 (dashed lines) with the same notations as in Fig.~\ref{fig:thermo}. Other results from the literature are also represented: NV~\cite{Negele:1973} (full circles), BSk24~\cite{Pearson18} (stars), and DHM~\cite{DouchinHaensel2000} (squares). See text for more details.}
\label{fig:neutronsCrust}
\end{figure*}

We now present more detailed results specific to our crust modeling that cannot be obtained from a phase coexistence model, as in Ref.~\cite{Gupta:2023}.
The number of neutrons in the gas $N_g$, in the nuclear cluster $N_\cl$, and the number of protons in the nuclear cluster $Z_\cl$ are shown in Fig.~\ref{fig:neutronsCrust}. We compare our results with other results obtained in the literature: the quantum calculation performed by Negele and Vautherin (NV)~\cite{Negele:1973} (full circles), the semi-classical calculation with quantum corrections based on BSk24 Skyrme-type interaction (BSk24)~\cite{Pearson18} (stars), and the CLDM construction by Douchin, Haensel, and Meyer (DHM)~\cite{DouchinHaensel2000} (squares). As previously noted, although our CLDM model does not contain quantum shell effects, it is performing quite well in predicting a rather constant value for the number of protons~\cite{Grams2022a,Grams2022c}. 

The number of neutrons in the gas phase is the quantity that is the most influenced by the improved modeling of dilute neutron matter. It is increased by up to $\approx 60$ for densities $n_B\approx 3~10^{-3}$~fm$^{-3}$, which represents a correction of 8-10\% to the prediction of the same quantity based on the mean field approximation. The difference between our results and others in the literature is much larger. These differences reflect the influence of the nuclear interaction itself as well as of the many-body correlations such as the shell effects. They have already been discussed, see for instance our recent analyses~\cite{Grams2021,Grams2022a,Grams2022b,Grams2022c}.

Note the qualitative good agreement between our results and DHM and BSk24 concerning $N_g$ and $N_\cl$, and we are close to BSk24 concerning $Z_\cl$. BSk24 results are computed with the semi-classical Extended-Thomas-Fermi approach with quantum corrections added perturbatively with the Strutinsky Integral.
The shell effects, which are not included in the present work, have been discussed in Ref.~\cite{Grams2022c}. The large differences observed with NV results mostly originate from the nuclear interaction employed in NV since at the moment of this work, advanced modeling of neutron matter as well as of exotic nuclei was unknown.

Finally, the authors of Ref.~\cite{Gupta:2023}  found an increase of the outer-inner crust transition density in the context of the phase coexistence model. Our prescription for the neutron correlations showed no difference in this transition density with respect to our original $H_2$ model: all models predict $n_{\rm drip} = 2.4 \times 10^{-4}$ fm$^{-3}$ consistently to the value obtained in Ref.~\cite{Grams2022a}. 
The outer-inner crust transition is obtained when the neutron chemical potential in the outer crust becomes positive, for which surface and Coulomb contributions to the energy are the most important ingredients. The impact of the correlations in the neutron gas is very weak since their appear at higher densities. 
The core-crust density transition is however more impacted by the neutron correlations. We obtained $n_{\rm cc} =  9.24 \times 10^{-2}$ fm$^{-3}$ for $H_2$ and, $n_{\rm cc}$ =  (9.06, 8.76, 8.33, 7.73, 7.12) $\times 10^{-2}$fm$^{-3}$, for ext1-ext5 respectively. The origin of the decrease of the transition density $n_{\rm cc}$ lies in the change of the proton fraction as a function of the strength of the neutron correlations: the larger the correlations, the more neutron-rich the crust, see Fig.~\ref{fig:thermo}. Once can also refer to the symmetry energy as a function of the neutron correlation shown in Fig.~\ref{fig:esym}. 
Since $n_{\rm drip}$ is unchanged and $n_{\rm cc}$ is reduced by the neutron correlations, the thickness of the NS inner-crust is reduced, which may impact its moment of inertia and NS oscillation modes and glitches.  

\section{Conclusions}
\label{sec:conclusions}

In this work, we discuss results for dilute neutron matter constrained by 'ab initio' predictions and we estimate the impact of the accurate description of dilute neutron matter for the properties of the crust of neutron stars. 

Dilute neutron matter, although strongly connected to the universal unitary limit, is not at the unitary limit. For instance, 'ab initio' predictions for the ratio $E_\NM/E_\ffg$ never reach the value given by the Bertsch constant for a universal unitary gas. There are several 'ab initio' predictions for this ratio showing that it is close to a value of about $0.6$ for densities $n_n\approx 10^{-4}$-$10^{-2}$~fm$^{-3}$. Note that despite the simplicity of the interaction, there are still some differences between the different 'ab initio' calculations. We suggest a formalism that separates the mean field contribution to one of the additional correlations. By adjusting the meta-model to the MBPT prediction in dilute neutron matter, the contribution from the mean-field can be calibrated, and we suggest a set of extensions that deviate from the mean-field predictions step-by-step. We label our extensions ext1 to ext5 by increasing integer reflecting the increasing differences with the mean-field reference. Our largest extensions ext4 and ext5 are comparable to the 'ab initio' predictions for dilute neutron matter. We then analyze the impact of our improved modeling in dilute neutron matter and find them quite substantial: up to about 30\% differences with the reference mean field prediction for some quantities and in a density region around $10^{-3}$~fm$^{-3}$.

We obtain a quantitative estimate of the impact of dilute neutron matter predictions for the physical system that exists in nature. Reversely, measured data related to the crust can be employed to test the modeling of the crust and its constituents, e.g., dilute neutron matter. To do so we apply our formalism to the CLDM description of cold-catalyzed neutron star crust and confront the effect of dilute neutron matter with the additional contributions from the nuclear clusters and the electrons. We obtain a large reduction of the effect of the correlations beyond mean-field in dilute neutron matter, which impacts the neutron star crust properties only up to 10\% or less.

We conclude our study relating dilute neutron matter and neutron star crust by remarking that the correlations beyond the mean field do not dominate the properties in the crust due to the additional contribution of other matter components: nuclear clusters and electrons. In other words, the important effects obtained in uniform matter finally appear to be small in the physical system where dilute neutron matter is present: the crust of neutron stars. Even if this effect is small, it is important to consider globally the dilute neutron matter constraint to the equation of state in the crust of neutron stars for both the mean field and the correlation part. The mean field was discussed in a previous analysis~\cite{Grams2022b}. In conclusion, dilute neutron matter remains important for quantitative calculations of the neutron star crust properties.

In the future, it would be interesting to study additional properties in the crust of neutron stars, such as the amount of super-fluid neutrons since it influences the cooling of neutron stars~\cite{Monrozeau07,Chamel13}, the neutron star seismology~\cite{Andersson21}, the neutron star glitches~\cite{Graber18} and possibly the release of fast radio bursts~\cite{Richard88,Qiao-Chu09}. It would also be interesting to study in more detail the interface between the crust and the core, the so-called pasta phase, as well as the effects of finite but low temperatures~\cite{Burrello22,Keller21} in the crust of neutron stars where correlations beyond the mean field may be quenched. The temperature at which correlations are suppressed would provide valuable information about their nature and their strength. The formalism presented here is well adapted (while it requires to be extended to finite temperature) for such studies, which are envisioned for future works.

\begin{acknowledgements}
We thank G. Col\'o, S. Gandolfi and I. Vida\~na for very interesting exchanges during the completion of this work. GG is supported by the Fonds de la Recherche Scientifique (F.R.S.-FNRS) and the Fonds Wetenschappelijk Onderzoek - Vlaanderen (FWO) under the EOS Projects nr O022818F and O000422F.
JM is supported by CNRS-IN2P3 MAC masterproject and benefits from PHAROS COST Action MP16214, as well as from the LABEX Lyon Institute of Origins (ANR-10-LABX-0066).
\end{acknowledgements}

Author contributions: G.G. performed most of the numerical analysis. All authors contributed to the preparation and revision of the manuscript.

Research data policy: The data created in this work and used for the figures of this manuscript is included as electronic supplementary material.

\bibliographystyle{spphys}       
\bibliography{biblio}

\end{document}